# Multistate Density Functional Theory for Local and Charge-Transfer Tripdoublet States from Triplet-Free Radical Interactions

Chenyu Liu,[1‡] Yang Xu,[2‡] Peng Bao,[3*] Yangyi Lu[2*] and Jiali Gao[1,2,4*]

1. Department of Chemistry and Supercomputing Institute, University of Minnesota, Minneapolis, MN 55455, USA
2. Institute of Systems and Physical Biology, Shenzhen Bay Laboratory, Shenzhen 518055, China
3. Beijing National Laboratory for Molecular Sciences, State Key Laboratory for Structural Chemistry of Unstable and Stable Species, Institute of Chemistry, Chinese Academy of Sciences, Beijing 100190, China
4. School of Chemical Biology and Biotechnology, Beijing University Shenzhen Graduate School, Shenzhen 518055, China

‡   Authors contributed equally.

*   Corresponding authors: Email, PB, baopeng@iccas.ac.cn; YL, luyy@szbl.ac.cn; JG, gao@jialigao.org

**Abstract**

The interaction between excited states of a closed-shell chromophore and a nearby free radical species gives rise to spin-coupled doublet states, namely singdoublet and tripdoublet, as well as a quartet state. This coupling facilitates transitions that are otherwise spin-forbidden, thereby enhancing intersystem crossing and influencing luminescence and non-radiative decay pathways. In this chapter, we explore these interactions using multistate density functional theory (MSDFT). By employing a minimal active space (MAS) comprising just ten determinant configurations, MSDFT effectively captures local and charge-transfer excitations with inclusion of correlation effects. MSDFT extends the Hohenberg-Kohn density functional theory from the ground state to encompass all electronic states, underscoring the potential for developing computationally efficient methods to study excited states. Numerical results demonstrate that MSDFT accurately reproduces both qualitative trends and quantitative excited-state energies, in accord with previous studies using extended multistate complete-active-space second-order perturbation theory (XMS-CASPT2). The work explores energy changes along a reaction path from the $D_0/D_1$ minimum energy crossing intersection to the $D_2/D_3$ crossing in the exciplex formed by 10-methylphenothiazine and a dicarboximide electron acceptor linked to the stable free radical 2,2,6,6-tetramethylpiperidin-1-oxyl (TEMPO).

# 1. Introduction

The precise control of electron spin in molecular systems has become increasingly critical for applications spanning organic light-emitting diodes, molecular spintronics, and quantum information science.[1-3] Early studies in porphyrin photophysics — notably by Gouterman and coworkers[4] — established an insightful framework for understanding these systems.[5] In particular, the four-orbital model elucidated the characteristic intense Soret and subtler Q bands observed in porphyrin spectra and highlighted how ligand-centered π→π* excitations (typically forming triplet states) can interact with a metal's unpaired d electron (a doublet state).[6,7] In metalloporphyrins, such as those containing copper(II) or vanadyl ions, this interaction produces mixed "tripdoublet" states,[8] where the overall spin remains a doublet, but the excitation is delocalized between the ligand and the metal center. This coupling not only "activates" transitions that would otherwise be spin-forbidden in a pure triplet but also underpins efficient intersystem crossing (EISC), thereby influencing luminescence and non-radiative decay pathways in these systems. The interplay between triplet and doublet states has gained renewed attention, and the unique spin properties of these systems are being harnessed in molecular spintronics.[9,10] Recent studies have expanded the scope of triplet-doublet interactions to include covalent chromophore-radical dyads, where strong exchange coupling between triplet and radical spins generates high-spin states such as quartets or quintets.[11,12]

In the present study, we consider the photochemical reaction (Scheme 1) between the triplet state of a closed-shell donor chromophore 10-methylphenothiazine (MPTZ) and an electron acceptor linked to the stable free-radical 2,2,6,6-tetramethylpiperidin-1-oxyl (**A-R$^\bullet$**).[13, 14] Formally, the reaction yields a biradical radical-ion pair (BRIP), which dissociates and can be detected as separate ions.

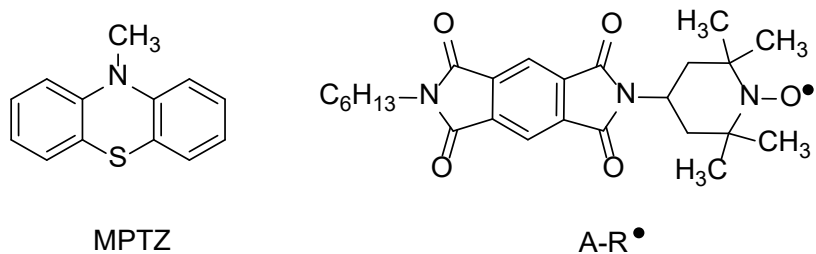

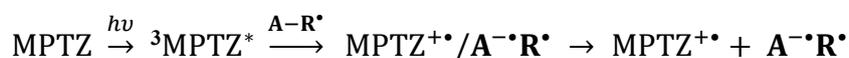

**Scheme 1**. Photochemical reaction to yield a biradical radical-ion pair (BRIP) between 10-methylphenothiazine (MPTZ) and the electron acceptor N-hexyl-benzene-1,2:3,4-(dicarboximide) linked to 2,2,6,6-tetramethylpeperidin-1-oxyl radical (**A-R$^\bullet$**).

It was found that there is a magnetic field effect on the photochemical reaction, increasing the free ion yield by 1.9-fold at a field strength of 2 T relative to that under zero external field. The process is closely related to two sets of doublet-quartet states, one involving the complex between a charge-transfer biradical (**A$^{-\bullet}$-R$^\bullet$**) and the radical ion (MPTZ$^{+\bullet}$), and the other that is characterized by the interaction of the local triplet ($T_1$) and singlet ($S_1$) states of MPTZ and the molecular radical doublet ground state ($D_0$) of **A-R$^\bullet$**. We use multistate density functional theory (MSDFT) to localize excited states to investigate the intricate spin coupling and charge transfer processes, which cannot be adequately treated by KS-DFT and linear-response TDDFT.

State-of-the-art computational approaches such as the extended multistate complete-active-space with second-order perturbation (XMS-CASPT2) now offer reliable means to model the complex excited-state dynamics of open-shell systems,[14-

[17] encompassing the interplay between ligand-based triplet states and metal-centered doublet states. For these systems, standard time-dependent density functional theory (TDDFT)[18] does not work since the excited states of open-shell systems in general involve double electron transitions which are not accounted for in linear response (LR) theories,[19] though methods to treat open-shell systems have been developed.[20-22] Furthermore, these methods utilize delocalized orbitals, which do not provide direct insights into interactions between excited states of individual fragments. A promising theoretical tool is multistate density functional theory presented in this article.[23-25] MSDFT can be used for addressing the challenges associated with closely spaced and interacting excited states, especially in systems where conventional single-state approaches suffer from spin contamination and a lack of state interaction.[26] By treating multiple states on an equal footing, MSDFT provides a balanced description of both static and dynamic correlation effects, making it particularly well-suited for exploring triplet–doublet interactions in open-shell materials and related magnetic molecules.[27,28] Thus, MSDFT can offer further insights into the subtle electronic factors that control intersystem crossing, luminescence, and the overall spin management strategies.

In the following, we first introduce the theoretical background and computational methods of multistate density functional theory. Then, we describe the computational details and optimization of minimal active space for different species. This is followed by results and discussion on molecular monomers and excited-state complexes. Finally, we summarize the main findings of the present study.

## 2. Theoretical Background

Recently, a series of fundamental theorems were proven in density functional theory,[29, 30] extending the Hohenberg-Kohn theory for the ground state to all electronic states.[31] These theorems rigorously establish MSDFT as a complete quantum theory. In this section, we introduce matrix density as the fundamental variable for excited states in density functional theory. This is followed by a summary of key concepts of MSDFT, and the definition of correlation matrix functional with respect to a minimum active space for $N$ electronic states.

### *2.1. Matrix Density*

Let $\hat{H} = \hat{H}^o + \sum_j^n v_{ext}(\boldsymbol{r}_j)$ be the Hamiltonian of an $n$-electron molecular system, where $\hat{H}^o$ consists of the kinetic and electronic repulsion operators and $v_{ext}(\boldsymbol{r})$ is the local external potential. Suppose that we were interested in the lowest $N$ eigenstates of the Hamiltonian, $\hat{H}\Psi_I = E_I \Psi_I$, where $I = 1, 2, \cdots, N$, and $E_1 \leq E_2 \leq \cdots \leq E_N$. The Hilbert subspace $\mathbb{S}_{min}^N$ spanned by these eigenstates is closed and invariant with respect to $\hat{H}$. Throughout this chapter, we use $I, J, \cdots$ to specify eigenstates (e.g., $\Psi_I$), $A, B, \cdots$ to denote basis states (typically multiconfigurational wave functions, $\Phi_A$), and lower case Greek letters $\zeta, \eta, \cdots$ to indicate determinant configurations (e.g., $\Xi_\eta$).

We first introduce the physical property, called $N$-matrix density, $\boldsymbol{D}(\boldsymbol{r})$ that is associated with an arbitrary set of orthonormal basis vectors $\{\Phi_A; A = 1, \cdots, N\}$ of $N$-dimensional Hilbert subspace $\mathbb{S}^N$:[32]

$$\boldsymbol{D}(\boldsymbol{r}) = \begin{pmatrix} D_{11}(\boldsymbol{r}) & \cdots & D_{N1}(\boldsymbol{r}) \\ \vdots & \ddots & \vdots \\ D_{1N}(\boldsymbol{r}) & \cdots & D_{NN}(\boldsymbol{r}) \end{pmatrix} \qquad (1)$$

where the matrix elements are given by $D_{AB}(r) =< \Phi_A|\hat{\rho}(r)|\Phi_B >$ with $\hat{\rho}(r)$ being the one-electron density operator $\hat{\rho}(r) = \sum_{j=1}^{n} \delta(r - r_j)$. In fact, any set of linearly independent states of $\mathbb{S}_{min}^N$ can be written in terms of its eigenstates $\{\Psi_I; I = 1, \cdots, N\}$:

$$\Phi_A = \sum_I^N u_{AI} \Psi_I; A = 1, \cdots, N$$

where $u_{AI}$ is an element of the unitary transformation matrix $\boldsymbol{U}$. The trace of the $N$-matrix density (Eq. 1) is invariant to a unitary transformation, thus, independent of the basis states of $\mathbb{S}_{min}^N$ used to represent $\boldsymbol{D}(r)$. We denote $\boldsymbol{D^o}(r)$ as the matrix density corresponding to the $N$ eigenstates, and the invariance property of the subspace density $\rho_v(r)$ is given by

$$\rho_v(r) \equiv \frac{1}{N} tr\{\boldsymbol{D^o}(r)\} = \frac{1}{N} tr\{\boldsymbol{D}(r)\} = \frac{1}{N} tr\{\boldsymbol{U^\dagger D^o}(r)\boldsymbol{U}\} \quad (2)$$

Reference 32 addressed the question of the required fundamental variable for a density functional theory of multiple electronic states. It was shown that the matrix variable $\boldsymbol{D}(r)$ is both necessary and sufficient to be used to resolve the energies of the individual eigenstates. In the special case of $N = 1$, $\mathbb{S}_{min}^N$ contains only the ground state of the system; thereby, the ground-state density $\rho_0(r)$ is sufficient to determine the ground-state energy, which is the well-known result of Hohenberg and Kohn.[31]

Apparently, it may come as a surprise, from referee comments and other contexts, that the off-diagonal elements of $\boldsymbol{D}(r)$, i.e., transition densities, are needed to resolve the energies of individual eigenstates, given that the trace is invariant and does not contain transition densities (Eq. 2). To fully answer this question, one needs to consider the Hamiltonian matrix functional (the first theorem of MSDFT below), whose trace is also invariant within $\mathbb{S}_{min}^N$ and is the quantity (not individual states) that is variationally minimized. The off-diagonal terms of the Hamiltonian matrix functional $\mathcal{H}[\boldsymbol{D}]$ are functionals of the transition densities, more precisely, the entire $N$-matrix density $\boldsymbol{D}(\boldsymbol{r})$. These elements determine the spectrum of eigenstates, that is, the relative energies of all different states in $\mathbb{S}_{min}^N$. In other words, it is the special set of densities (diagonal) and transition densities (off-diagonal) comprising $\boldsymbol{D}^o(\boldsymbol{r})$ that yields a Hamiltonian matrix functional $\mathcal{H}[\boldsymbol{D}^o]$ in diagonal form. For any $N$-matrix density $\boldsymbol{D}(\boldsymbol{r})$ other than $\boldsymbol{D}^o$, the corresponding Hamiltonian matrix functional $\mathcal{H}[\boldsymbol{D}]$ possesses non-zero off-diagonal elements that dictate state interactions. Notably, the trace of $\mathcal{H}[\boldsymbol{D}]$ remains invariant regardless of these off-diagonal terms. $\boldsymbol{D}^o(\boldsymbol{r})$ and its corresponding eigenenergies can be obtained by diagonalizing $\mathcal{H}[\boldsymbol{D}]$, whether the exact or an approximate matrix functional is used. Without the complete Hamiltonian matrix functional, the energy derived from a functional of a non-ground-state density is not guaranteed to correspond to an eigenstate of the Hamiltonian $\widehat{H}$, as there is generally no variational principle applicable to excited states.

It is also of interest to note that the transition dipole moment $d_{IJ}$ between states $I$ and $J$ can be determined by integration of the corresponding transition density, i.e., an off-diagonal element of $\boldsymbol{D}^o(\boldsymbol{r})$: $d_{IJ} = e\int \boldsymbol{r}D_{IJ}^o(\boldsymbol{r})d\boldsymbol{r}$. This further highlights the critical role of transition densities in determining the spectrum of excited states. Figure 1 depicts the matrix density of a hydrogen molecule at its equilibrium geometry, including the three lowest singlet states. Considering the symmetry properties of transition densities and the spatial coordinates $r = (x, y, z)$, the excitation from the ground state $S_0(\Sigma_g^+)$ to the first excited state $S_1(\Sigma_u^+)$ possesses a non-zero transition dipole moment. In contrast, the transition dipole moment to the second excited state $S_2(\Sigma_u^-)$ is zero.

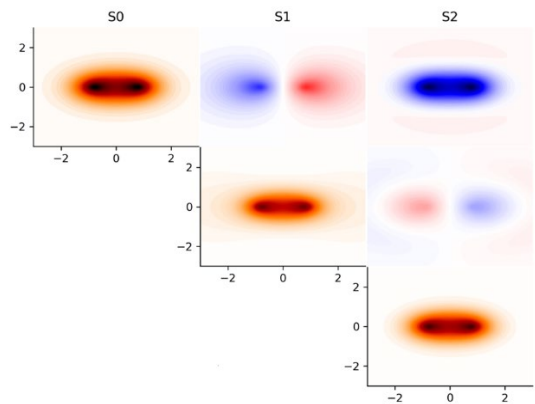

Figure 1. Illustration of the matrix density of the first three singlet states of hydrogen molecule at its equilibrium geometry, $R(\text{H} - \text{H}) = 1.40$ Bohr).

*2.2. Multistate Density Functional Theory*

Multistate Density Functional Theory (MSDFT) is predicated upon a series of fundamental theorems. While the exact formulations of these theorems are detailed in reference 23, we will summarize their main results here.

i. By introducing the matrix density of rank $N$ (Eq. 1) as the fundamental variable, Theorem 1 of MSDFT proves that there is a one-to-one correspondence between $D(r)$ and the Hamiltonian matrix $\mathcal{H}$ within the subspace $\mathbb{S}_{min}^N$.[23] This allows the Hamiltonian matrix to be expressed as a *matrix functional* of $\boldsymbol{D}(\boldsymbol{r})$:

$$\boldsymbol{\mathcal{H}}[\boldsymbol{D}] = \boldsymbol{\mathcal{F}}[\boldsymbol{D}] + \int \boldsymbol{D}(\boldsymbol{r})v_{ext}(\boldsymbol{r})d\boldsymbol{r} \qquad (3)$$

where $\boldsymbol{\mathcal{F}}[\boldsymbol{D}]$ is the universal matrix functional, independent of the external potential $v_{ext}(\boldsymbol{r})$. The existence of the universal matrix functional $\boldsymbol{\mathcal{F}}[\boldsymbol{D}]$ was subsequently proven recently.[33]

ii. Theorem 2 establishes a variational principle for determining the total energies of eigenstates within the subspace $\mathbb{S}_{min}^N$. Specifically, variational optimization of the trace of the Hamiltonian matrix, called multistate energy $E_{MS}[\boldsymbol{D}] = tr\{\boldsymbol{\mathcal{H}}[\boldsymbol{D}]\}$, enables simultaneous determination of the energies and electron densities of all $N$ eigenstates. Therefore, given a trial density $\boldsymbol{D}'(\boldsymbol{r})$, the variation principle is given as follows:[23]

$$E_{MS}[\boldsymbol{D}'] \geq E_{MS}[\boldsymbol{D}] \qquad (4)$$

where the equal sign holds true if the trial density is identical to that of the subspace $\mathbb{S}_{min}^N$.

iii. Theorem 3 introduces an algorithm for representing $N$-matrix densities and, notably, establishes an upper bound on the associated computational cost. It states that an $N$-matrix density $\boldsymbol{D}(\boldsymbol{r})$ can be exactly mapped to a set of $N$ multiconfigurational wave functions, $\{\Phi_A; A = 1,\cdots, N\}$, which can be expressed as linear combinations of no more than $N^2$ independent (non-orthogonal) Slater determinants.[23, 29] An important consequence is that a minimal active space (MAS) can be formulated with a computational scaling of $O(N^2)$ if $N$ eigenstates are being concerned.[23, 29]

MSDFT extends the traditional Hohenberg-Kohn theorems, originally formulated for ground-state systems,[31] to encompass excited states by minimizing a multistate energy functional. This approach addresses the limitations of conventional density functional theory,[21, 34] which relies on linear response methods to compute excited-state energies.[35] MSDFT employing nonorthogonal state interaction exemplifies a "dynamic-then-static" approach,[36] wherein dynamic electron correlation is first incorporated during the optimization of individual configurations, followed by configuration interaction to account for static correlation effects.[24, 25] An analogy in wave function theory is the extended multistate complete-active-space second-order perturbation theory (XMS-CASPT2),[14-16] though it is a static-dynamic-then-static procedure.

*2.3. Minimal Active Space.*

Theorem 3 of MSDFT[23] noted above defines a minimal active space (MAS), $V_{MAS} = \{\Xi_\xi; \xi = 1, \cdots, M\}$,[29, 37] consisting of no more than $N^2$ Slater determinants, i.e., $M \leq N^2$. Then, a set of auxiliary multiconfigurational wave functions can be constructed to exactly represent $D(r)$.[29]

$$\Phi_A = \sum_\xi^M C_{\xi A} \Xi_\xi \tag{5}$$

where $A = 1, \cdots, N$, $\Xi_\xi$ is the $\xi th$ Slater determinant constructed from $n$ (the number of electrons) one-body spin orbitals $\{\psi_{i\sigma}^\xi\}$, and $\{C_{\xi A}; \xi = 1, \ldots, M\}$ are the configuration coefficients for auxiliary state $A$. As usual, each molecular orbital is expressed as a linear combination of atomic basis functions. In general, the orbitals of the same determinant are orthonormal, $<\psi_{i\sigma_i}^{\xi}|\psi_{j\sigma_j}^{\xi}> = \delta_{ij}\delta_{\sigma_i\sigma_j}$, but they are nonorthogonal from different determinants.

Both the orbital and configuration coefficients (Eq. 5) can be optimized simultaneously in a nonorthogonal self-consistent-field (NOSCF) manner.[29] Alternatively, one can optimize the orbitals in each determinant state first, followed by a single step of nonorthogonal state interaction (NOSI),[29] which is the approach adopted in the present study. Then, each element of the $N$-matrix density $\boldsymbol{D}(\boldsymbol{r})$ is determined as follows:

$$D_{AB}(r) = <\Phi_A|\hat{\rho}(r)|\Phi_B> \qquad (6)$$

It is of interest to note that any existing nonorthogonal MCSCF or NOCI quantum chemistry program can be adopted for MSDFT-NOSCF or NOSI calculations. NOSI distinguishes from NOCI in that dynamic correlation is included as a functional in determinant configurations in the first place.[25]

We comment that the auxiliary wave functions in Eq. (5) are used purely for the purpose of representing the matrix density $\boldsymbol{D}(\boldsymbol{r})$, just as the use of a single KS-determinant to represent the ground-state density.[32] One does not have to implicate further meanings to the multiconfigurational auxiliary wave functions, although computational results indicate that Eq. (5) is also a good approximation to the many-body wave functions at a quality similar to or better than those from MS-MCSCF methods.[38]

*2.4. Correlation Matrix Functional.*

The minimal active space $V_{MAS}$ for the exact mapping of $D(r)$ naturally leads to the definition of a correlation matrix functional,[30, 33] making use of the universal matrix functional introduced in Eq. (3):

$$\mathcal{E}_c[D] = \mathcal{F}[D] - T_{ms}[\{\Phi_A\}] - E_{Hx}[\{\Phi_A\}] \tag{7}$$

where the arguments of the last two terms remind the fact that they are obtained from the auxiliary wave functions in Eq. (5). They are determined as in standard multiconfigurational wave function calculations as follows:[24]

$$H^o_{AB} = <\Phi_A|\widehat{H}^o|\Phi_B> = (T_{ms})_{AB} + (E_{Hx})_{AB} \tag{8}$$

Then, the total Hamiltonian matrix density functional in auxiliary state basis (Eq. 5) is explicitly given by

$$\mathcal{H}_{AB}[D] = <\Phi_A|\widehat{H}|\Phi_B> + (\mathcal{E}_c[D])_{AB} \tag{9}$$

where $<\Phi_A|\widehat{H}|\Phi_B> = H^o_{AB} + \int D_{AB}(r)v_{ext}(r)dr$. The notation $(\mathcal{E}_c[D])_{AB}$ emphasizes that the matrix element $(\mathcal{E}_c)_{AB}$ is generally dependent on the full $N$-matrix density $D(r)$.

Recall that Eq. (7) represents an extension of the definition of the exchange-correlation functional in Kohn-Sham density functional approach for the ground state[39] to a collection of the lowest $N$ eigenstates with respect to the MAS $V_{MAS}$.[30, 33] Unlike the ground state, there is no rigorous definition of electron correlation for an excited state, but the Hamiltonian matrix density functional in Eq. (3), through state interaction, distributes the total correlation of the entire subspace in Eq. (7) to individual eigenstates in $\mathbb{S}_{min}^N$. The physical meaning of the correlation matrix functional is that it represents the remaining correlation effects outside of the minimal active space $V_{MAS}$.[25, 30]

**3. Method and Procedure**

The practical computation using MSDFT can be carried out self-consistently by optimizing both the orbital and configuration coefficients (Eq. 5) or using a nonorthogonal state interaction (NOSI) algorithm.[29] In the latter case, which is the method used in this study, we first optimize the orbitals of each determinant configuration in the MAS, followed by a single step of diagonalization of the Hamiltonian matrix functional. In this section, we describe the specific steps of the NOSI method and approximations for constructing the Hamiltonian matrix functional.

*3.1. Nonorthogonal State Interaction*

The computational procedure of nonorthogonal state interaction (NOSI) is closely related to the well-known nonorthogonal configuration interaction (NOCI) in wave function theory,[40] except that dynamic correlation is included in orbital optimization for each determinant configuration. NOSI was first used in the work of reference 24, and it is adopted in this study to determine the energies of ground and excited states. In NOSI and the following discussion, it is convenient to express the total Hamiltonian and correlation matrix functionals in terms of determinant configurations. Then, the Hamiltonian matrix density functional on a determinant basis is explicitly given by

$$H_{\xi\eta} = <\Xi_\xi|\hat{H}|\Xi_\eta> + (E_c)_{\xi\eta} \tag{10}$$

where $\xi$ and $\eta$ indicate determinant $\Xi_\xi$ and $\Xi_\eta$. Note that we have switched to normal Roman scripts to denote matrix functionals in a determinant basis.

The NOSI computational procedure involves two main steps:

1. **Optimization of Individual Determinants**: Each determinant $\Xi_\xi$ ($\xi = 1, \cdots, M$), representing ground or excited configurations within the active space, is optimized separately (see below). The number of determinants ($M$) in the minimal active space (MAS) may vary and is not strictly limited to $N^2$, where $N$ is the number of states of interest, as it depends on the inclusion of all spin-complementary configurations and diabatic states under consideration. Because individual excited states are optimized, it limits the application of NOSI to low-lying excited states where unoccupied orbitals contributing to these states have adequate separations from higher virtual orbitals.

2. **Multistate Interaction**: After obtaining the optimized determinants, the energies and auxiliary wave functions (or densities) of the individual eigenstates are determined by diagonalizing the Hamiltonian matrix functional spanned by the determinant basis, $\mathbf{HC} = \mathbf{SCE}$.[24] This step allows for the calculation of eigenstate energies and provides insights into the electronic structure of the system such as local and charge-transfer interactions.[41] Therefore, the energy functional for an individual state $I$ is given below.

$$E_I[\mathbf{D}] = \sum_\xi^M C_{\xi I}^2 H_{\xi\xi}(D_{\xi\xi}) + \frac{1}{2}\sum_{\xi \neq \eta}^M C_{\xi I} C_{\eta I} H_{\xi\eta} \qquad (11)$$

where $\{C_{\xi I}\}$ are configuration coefficients, and $\{H_{\xi\eta}\}$ are elements of the Hamiltonian matrix functional. In NOSI, the off-diagonal element $H_{\xi\eta}$ ($\xi \neq \eta$) generally depends on the state densities ($D_{\xi\xi}$) and determinant wave functions ($\Xi_\xi$) as well as spin-coupling (sc) interaction terms (($E_c^{sc})_{\xi\eta}$) and transition densities ($D_{\xi\eta}$).

*3.2. Diagonal terms of the Hamiltonian matrix functional*

Clearly, the key in practice is to develop approximations for the Hamiltonian matrix functional coupled with a method to optimize the orbitals for representing the matrix density. In general, there is no term-by-term correspondence in a matrix functional, e.g., $D_{\xi\eta} \leftrightarrow (E_c[D_{\xi\eta}])_{\xi\eta}$.[29] Rather, each element is a functional of the entire matrix function $(E_c[\boldsymbol{D}])_{\xi\eta}$. However, if we make the approximation $(E_c[\boldsymbol{D}])_{\xi\xi} \approx E_c[\rho_{\xi\xi}]$ for the diagonal elements of the correlation matrix functional,[24] the energy expression for these terms in Eq. 10 ($\xi = \eta$) becomes identical to that of KS-DFT using determinant $\Xi_\xi$. This suggests that we may adopt one of the exchange-correlation functionals developed for Kohn-Sham theory to determine $H_{\xi\xi}$.[24]

We determine the energies of the diagonal elements of the Hamiltonian matrix functional by performing occupation-constrained KS-DFT optimizations, wherein specific non-aufbau occupied determinants are enforced,[42, 43] except for the ground state.

$$H_{\xi\xi}[\boldsymbol{D}] = E_\xi^{KS}[\rho_\xi] = \langle \Xi_\xi|\hat{H}|\Xi_\xi\rangle + E_c^{KS}[\rho_\xi] \qquad (12)$$

where $E_\xi^{KS}[\rho_\xi]$ is the KS-DFT energy for determinant configuration $\Xi_\xi$, and $\rho_\xi(r)$ is the density determined from the occupied orbitals that form $\Xi_\xi$. Here, we may define the *Kohn-Sham correlation energy* as the difference between the KS-DFT energy and the Hartree-Fock energy using KS orbitals, $E_c^{KS}[\rho_\xi] = E_\xi^{KS} - E_\xi^{HF}$.

It is crucial to maintain a given non-aufbau occupation during optimization, and two methods have been developed in our group.[42-44] In this work, we employ the block-localized excitation (BLE) approach,[42] which uses a ground-state orbital projection to preserve the order of orbitals, $(T_0^\dagger F T_0)T = (T_0^\dagger S T_0)T\epsilon^{\mathrm{BLE}}$, where $T_0$ and $T$ are, respectively, the orbital coefficient matrices of the ground-state and the non-aufbau state. With this projection, the orbitals are expressed in the same order as the ground state during each self-consistent field (SCF) iteration, ensuring consistency in the non-aufbau state.

*3.3. Spin-multiplet degeneracy constraint for spin-coupling interactions*

The off-diagonal terms of the Hamiltonian matrix functional for spin coupling interactions among unpaired electrons can be determined consistently with respect to KS-DFT calculations for the highest-spin state for which a single determinant is adequate for the all spin-up configuration.[26] In particular, the spin pair interactions between two unpaired electrons yield a singlet state and a triplet state with $M_s = 0$. The latter state is degenerate with the all spin-up configuration ($M_s = 1$). Then, the element for the correlation matrix functional, $(E_c)_{\xi\eta}^{ST}$, between the two determinant configurations $\Xi_\xi^{\uparrow\downarrow}$ and $\Xi_\eta^{\downarrow\uparrow}$ can be rigorously determined by

$$(E_c)_{\xi\eta}^{ST} = E_c^{KS}[\rho_{\xi_T}^{\uparrow\uparrow}] - E_c^{KS}[\rho_{\xi\xi}^{\uparrow\downarrow}] \qquad (13)$$

where $E_c^{KS}[\rho_{\xi_T}^{\uparrow\uparrow}(\mathbf{r})]$ is the KS correlation energy of the spin-up triplet state $\Xi_{\xi_T}^{\uparrow\uparrow} = |\psi_{1\alpha}^\xi \psi_{1\beta}^\xi \cdots \psi_{(n/2-1)\beta}^\xi \boldsymbol{\psi}_{i\alpha}^\xi \boldsymbol{\psi}_{j\alpha}^\xi >$, and $E_c^{KS}[\rho_{\xi\xi}^{\uparrow\downarrow}]$ is the KS correlation energy of the spin-mixed determinant state $\Xi_{\xi\xi}^{\uparrow\downarrow} = |\psi_{1\alpha}^\xi \psi_{1\beta}^\xi \cdots \psi_{(n/2-1)\beta}^\xi \boldsymbol{\psi}_{i\alpha}^\xi \boldsymbol{\psi}_{j\beta}^\xi >$. Of course, the spin-flip determinant $\Xi_{\eta\eta}^{\downarrow\uparrow} = |\psi_{1\alpha}^\eta \psi_{1\beta}^\eta \cdots \psi_{(n/2-1)\beta}^\eta \boldsymbol{\psi}_{i\beta}^\eta \boldsymbol{\psi}_{j\alpha}^\eta >$ has an identical energy and can be equivalently used. The use of Eq. (13) to determine the off-diagonal element of the correlation matrix functional between spin complementary states implies that an extra calculation, for the all spin-up configuration $\Xi_{\xi_T}^{\uparrow\uparrow}$, is needed. Nevertheless, the increased computational cost is negligible.

In the case of three electrons in three orbitals (Figure 2), critical in the preset study, a similar relationship can be derived by considering the transition from the highest doubly occupied orbital (H) to the lowest unoccupied orbital (L) of a free radical system in which the unpaired electron is in the singly occupied orbital (S).[27, 28] To enforce the energy degeneracy between the $M_S = 1/2$ and $M_S = 3/2$ multiplets of the quartet state $(S = 3/2)$ from MSDFT, we compare the conditions in multiconfigurational wave function theory which yields two doublet states and a quartet state with $M_S = 1/2$. The latter has an identical energy as that of the single determinant configuration $\Phi_4$. According to the Slater-Condon rules, the energy difference between the highest-spin $(S + 1 = 3/2)$ determinant $\Phi_4$ and the lower-spin $(S = 1/2)$ determinants $(\Phi_1 \sim \Phi_3)$ can be expressed in terms of the exchange integrals $K_{ij}$ among the three orbitals $(i, j = L, S,$ and $H)$ in Figure 2.

$$\begin{pmatrix} \Delta E_{41}^{HF} \\ \Delta E_{42}^{HF} \\ \Delta E_{43}^{HF} \end{pmatrix} = \begin{pmatrix} 1 & 0 & 1 \\ 0 & 1 & 1 \\ 1 & 1 & 0 \end{pmatrix}^T \begin{pmatrix} K_{LS} \\ K_{LH} \\ K_{SH} \end{pmatrix} \qquad (14)$$

where $K_{ij}$ ($i,j = L, S, H$) are the exchange integrals for two electrons in orbitals $i$ and $j$, and $\Delta E_{4A}^{HF} = E^{HF}(\Phi_4) - E^{HF}(\Phi_A)$ are the Hartree-Fock energy difference between configurations $\Phi_4$ and $\Phi_A$ with $A = 1, 2, 3$ (Figure 2). It is the linearity relationship in Eq. (14) that ensures energy degeneracy of spin multiplets. In multistate density functional theory, one must respect this symmetry condition albeit using an exchange-correlation functional that breaks such symmetry.[26, 27]

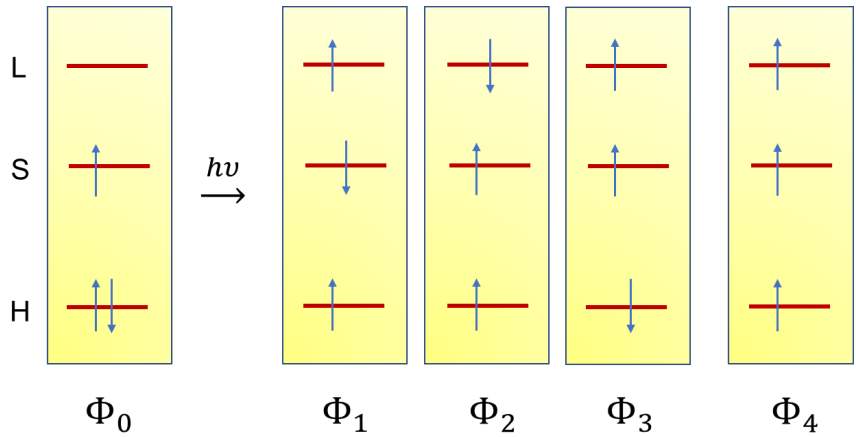

**Figure 2**. Determinant configurations of the doublet and quartet spin complementary states with three electrons in three orbitals. Configurations with $M_S > 0$ are shown.

Now that these exchange integrals correspond to the resonance integrals between two determinants in MSDFT, we change the orbital indices for the exchange integrals in Eq. (14) to the corresponding determinants that differ by flipping the spins of the two electrons in the two orbitals. Then, we obtain the correspondence: $K_{LS} \sim H_{12}^{\text{MSDFT}}$, $K_{LH} \sim H_{13}^{\text{MSDFT}}$, and $K_{SH} \sim H_{23}^{\text{MSDFT}}$. Thus, the elements of the Hamiltonian matrix functional that restore spin symmetry to match the energy degeneracy of the multiplets of the quartet, $E_Q(|\frac{3}{2},\frac{1}{2}>) = E_4^{KS}(\Phi_4)$, are given as follows:

$$\begin{pmatrix} H_{12}^{MSDFT} \\ H_{13}^{MSDFT} \\ H_{23}^{MSDFT} \end{pmatrix} = \frac{1}{2} \begin{pmatrix} 1 & -1 & 1 \\ -1 & 1 & 1 \\ 1 & 1 & -1 \end{pmatrix} \begin{pmatrix} \Delta E_{41}^{KS} \\ \Delta E_{42}^{KS} \\ \Delta E_{43}^{KS} \end{pmatrix} \qquad (15)$$

where $\Delta E_{4A}^{KS} = E^{KS}(\Phi_4) - E^{KS}(\Phi_A)$ are the energy difference from Kohn-Sham DFT calculations, with $A = 1, 2,$ and 3. Notice that we have directly used the inverse of the matrix transpose in Eq. (14).[27]

*3.4. Other off-diagonal elements of correlation matrix functional*

The off-diagonal elements of $E_c[\mathbf{D}]$ for all other situations are determined by the overlap-weighted average correlation energy of the two interacting states, which is the leading term to ensure subspace invariance symmetry:[24, 30]

$$(E_c)_{\xi\eta} = \frac{1}{2} S_{\xi\eta} \{ E_c^{KS}[\rho_\xi] + E_c^{KS}[\rho_\eta] \} \qquad (16)$$

In Eq. (16), $S_{\xi\eta}$ is the overlap integral between determinants $\Xi_\xi$ and $\Xi_\eta$.

**4. Computational Details**

For the closed-shell monomer 10-methylphenothiazine (MPTZ), the MAS includes nine determinant configurations: the Kohn-Sham determinant for the ground state along with those singly excited configurations from the two highest occupied and two lowest unoccupied orbitals. We found that this is sufficient for determining the ground and the first two singlet excited states and two triplet states. For the dicarboximide electron acceptor linked to 2,2,6,6-tetramethylpiperidin-1-oxyl **A-R**• (Scheme 1), we employ a MAS consisting of six determinant configurations ($\Phi_0$ through $\Phi_3$ in Figure 2,[27, 28] plus the $H \rightarrow S$ and the $S \rightarrow L$ excited determinants along with the Kohn-Sham determinant). This yields four doubly excited states and one quartet state in addition to the ground state doublet radical.

For the donor (MPTZ) and acceptor (**A-R**•) exciplex, we consider both local and charge transfer excitations. To describe these effective diabatic states, we first partition the molecular complex into two molecular fragment blocks, in which molecular orbitals are strictly block-localized (BLMO) by construction.[42, 45] Specifically, the BLMOs are expressed in terms of atomic orbitals situated on atoms of a given fragment, MPTZ or **A-R**• (Figures 3 and 4).

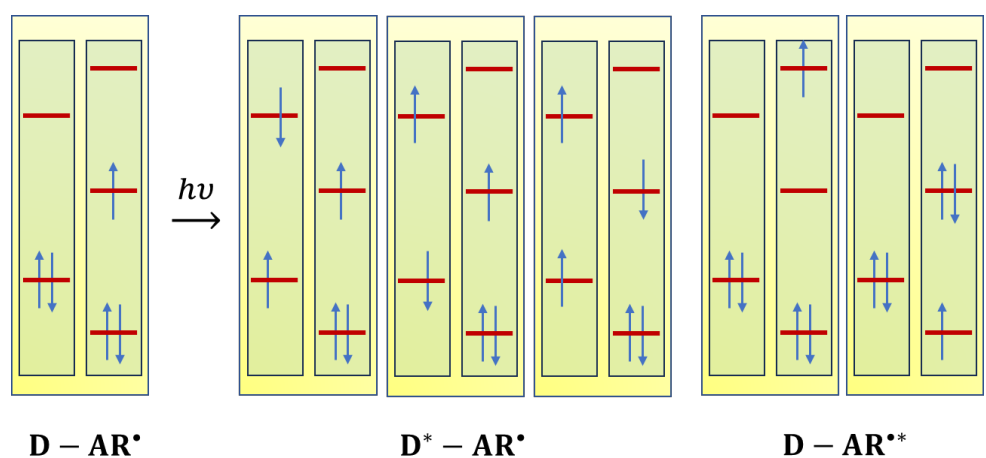

**Figure 3**. Illustration of block localization of molecular orbitals (BLMO) in the bimolecular complex MPTZ(**D**) and **A-R**•, both for the ground state **D-AR**•, MPTZ-locally excited $D^*$-**AR**•, and **A-R**•-locally excited $D$-**AR**•* configurations. Each fragment, **D** and **A-R**•, along with its block-localized frontier orbitals is colored in green, and the bimolecular complex in each electronic configuration is grouped into a box in yellow background, which is fully antisymmetrized as a block-localized determinant.

In the first group of locally excited states, the BLMOs and excited configurations are illustrated in Figure 3, corresponding to the interactions between singlet and triplet excited states of MPTZ with the doublet ground state of the receptor molecule **A-R**•, denoted by $D^*$-**AR**•, and between MPTZ in the ground state interacting with $H \rightarrow S$ and $S \to L$ excited configurations, given as **D-AR$^{•*}$**. We have also considered the $H \to L$ multiplets, which yield two doublet and one quartet states of **A-R$^{•}$** in the locally excited complex, but these states are higher in energy than states depicted in Figure 3 and the charge-transfer states in Figure 4 below. Thus, they are not included in further discussion.

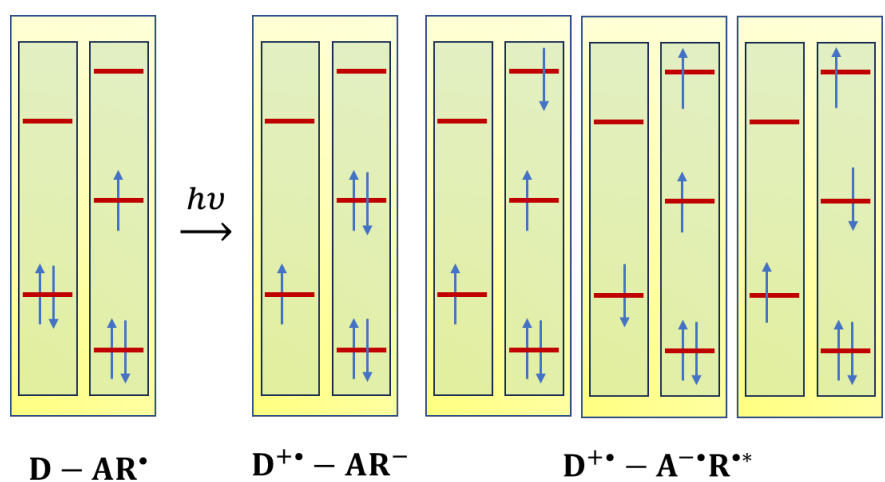

**Figure 4**. Charge transfer excited configurations giving rise to the radical cation and anion ion-pair single-determinant doublet state **D$^{+•}$-AR$^{-}$**, and the multiconfigurational biradical (anion) radical (cation) ion pair (BRIP) **D$^{+•}$-A$^{-•}$R$^{•*}$**. The latter includes two doublet states (singdoublet and tripdoublet) and one quartet state with $M_S = 1/2$.

The charge-transfer states considered in this study are shown in Figure 4. The direct forward electron transfer from the highest occupied orbital of MPTZ to the singly occupied orbital of the acceptor **A - R$^{•}$** yields an ion pair doublet state, corresponding to a singlet determinant configuration of **AR$^{-}$** in the complex **D$^{+•}$-AR$^{-}$**. Charge-transfer excitation to the lowest unoccupied orbital of **A-R$^{•}$** produces three possible spin arrangements (positive $M_S = 1/2$), resulting in two doublet states, called singdoublet (high energy) and tripdoublet (low energy), and one quartet state, all of which are of multiconfigurational character. One of the configurations in Figure 4 (also in Figure 3) involves double excitation, corresponding to one electron excitation plus a spin flip, which is not accessible by standard LR-TDDFT.[19] Spin-adapted TDDFT approaches have been developed and can be used to model excited states of open-shell molecules.[20, 21, 27]

In all, the ten unique electronic configurations shown in Figures 3 and 4 are used to form the minimal active space to describe the relevant excited states of the MPTZ-AR• exciplex.

Throughout this study, we used the hybrid PBE0 functional,[46, 47] which incorporates 25% of Hartree-Fock exchange into the PBE functional,[48] along with the cc-pVDZ basis set[49] for individual determinant optimizations and for constructing the Hamiltonian matrix functional. For the closed-shell MPTZ we found that the computed excitation energies from TDDFT calculations with a larger basis of aug-cc-pVTZ are within 0.1 eV of those from the double-zeta basis. In view of the size of the system, the basis set used in MSDFT is reasonable.

Monomer geometries in the ground and the first singlet excited state of MPTZ were optimized using, respectively, PBE0/cc-pVDZ and TDDFT/PBE0/cc-pVDZ.[50] For dimer structures, we took the coordinates reported in the work by Song,[14] who located the $D_0/D_1$ and $D_2/D_3$ minimum energy crossing intersection (MECI) and constructed a series intermediate structures interpolated along a pathway between the two crossing points. These MECI geometries (Figure 5) were optimized using state-averaged complete-active-space self-consistent-field (SA-CASSCF) with an active space of three electrons in four orbitals along with the polarizable continuum model (PCM) corresponding to methanol solvent. Song used these structures to perform XMS-CASPT2 calculations in a spin-free implementation.[14] The energies at the crossing points are no longer degenerate after dynamic correlation is corrected. In the present study, we did not include solvent effects in MSDFT calculations, making the energies of ion-pair states shift upwards. We analyzed solvation effects on the neutral and ion-pair states of the bimolecular complex along the interpolated pathway with PCM using Gaussian16.[50] For the ionic state, we protonated MPTZ at the sulfur site to create an ion-pair model to give a rough estimate of solvation energy. This enables the energy curves and trends to be compared with the XMS-CASPT2 results.

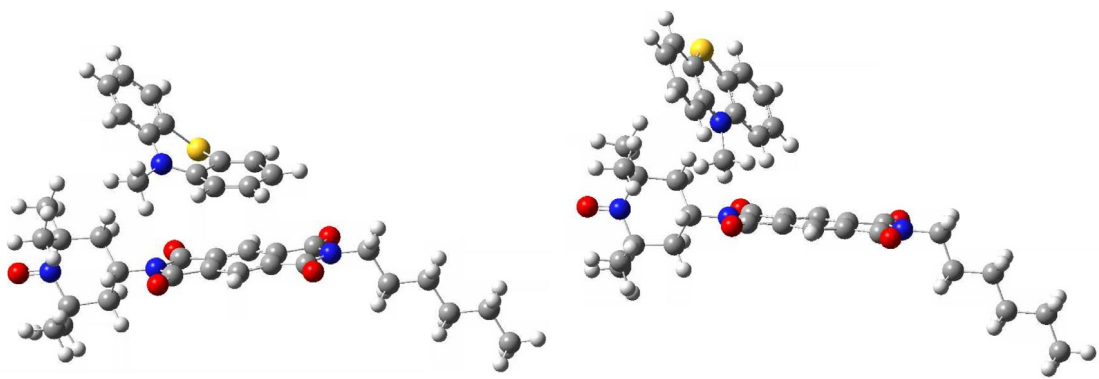

**Figure 5**. Geometries at the $D_2/D_3$ (left) and $D_0/D_1$ (right) minimum energy crossing intersection from SA-CASSCF[3e,4o]/6-31G(d) optimization along with a polarizable continuum model (PCM) from Reference [14]. These structures and several intermediate geometries interpolated along a path connecting these two points are used in this study.

## 5. Results and Discussion

### 5.1. Monomers.

Table 1 and Table 2 list the computed energies of the ground and excited states of 10-methylphenothiazine (MPTZ) and the radical species 2,2,6,6-tetramethylpiperidin-1-oxyl (TEMPO), the electron acceptor **A** linked to TEMPO, **A-R**$^\bullet$, and MPTZ$^{+\bullet}$. Both MSDFT and TDDFT results are listed for comparison and several geometries were used, including the optimized structures in the ground states ($S_0$ for closed-shell and $D_0$ for radical species) and the first singlet excited ($S_1$) state of MPTZ along with two monomer structures extracted from the complexes at the $D_2/D_3$ and $D_0/D_1$ MECI points. For the radical species, we only included excitation energies corresponding to the $H \to S$ single electron excitation from LR-TDDFT calculations because higher states are not adequately treated in these calculations. This comparison aims to demonstrate that MSDFT not only produces excitation energies closely aligning with those obtained from LR-TDDFT for molecules where the latter method is effective but also extends its applicability to open-shell molecules, for which LR-TDDFT is not directly applicable. Further, comparisons with experimental observations are made in the text.

**Table 1**. Computed ground and excited state energies (eV) 10-methylphenothiazine (MPTZ) using nonorthogonal state interaction of multistate density functional theory (MSDFT-NOSI) and time-dependent density functional theory (TDDFT) with PBE0 functional and cc-pVDZ basis set. Optimized geometries at the ground ($S_0$) and the first singlet excited ($S_1$) states are used along with two monomer structures extracted from the $D_2/D_3$ and $D_0/D_1$ minimum energy crossing interaction points of the complex, corresponding to frames 1 and 100 in Reference 14. All energies are given relative to the ground state of MPTZ, and values in parentheses are computed using the aug-cc-pVTZ basis set.

|  | MSDFT-NOSI | LR-TDDFT |
|---|---|---|

| Optimized Geometry | $S_0$ | $S_1$ | $S_2$ | $T_1$ | $S_1$ | $S_2$ |
|---|---|---|---|---|---|---|
| $S_0$ | 0.0 | 3.78 | 4.19 | 3.49 | 3.88 (3.79) | 4.26 (4.09) |
| $S_1$ | 0.41 | 3.13 | 4.23 | 2.79 | 3.25 | 4.23 |
| $D_2/D_3$ | 1.00 | 4.27 | 5.11 | 3.38 | 4.37 | 5.07 |
| $D_0/D_1$ | 0.86 | 3.46 | 4.47 | 3.12 | 3.61 | 4.47 |

It was reported that the $S_1$ excited state energy of MPTZ is 3.39 eV above its ground state experimentally.[1, 51] For comparison, the computed vertical excitation energy is 3.78 eV from MSDFT, in close agreement with TDDFT results (3.88 eV), but differing from the experimental value. However, the adiabatic excitation energies of the $S_1$ state from both MSDFT and TDDFT (3.13 and 3.25 eV) are in reasonable accord with experiments, based on TDDFT geometry. The emission energy from the $S_1$ state was reported to be 2.75 eV,[51] in excellent accord both with the MSDFT (2.72 eV) and TDDFT (2.84 eV) results (Table 1). We have separately optimized the triplet state geometry using PBE0, which is 2.82 eV (2.76 eV from MSDFT) above the $S_0$ ground state. This may be compared with an experimental value of 2.64 eV from charge-transfer experiments.[1, 51] For the two distorted structures in the exciplexes with **A-R•**, we find that the agreement between MSDFT and TDDFT is also good. In general, TDDFT performs very well for excitation energies on small organic molecules such as MPTZ.[52] The agreement between MSDFT and TDDFT shown in Table 1, as well as the accord with experimental data, suggest that the minimum active space used and the present NOSI method are adequate for modeling the first few excited states of MPTZ in the this study.

**Table 2**. Computed ground and excited state energies (eV) the cation radical of 10-methylphenothiazine (MPTZ$^{+\bullet}$), 2,2,6,6-tetramethylpiperidin-1-oxyl (TEMPO), and **A-R$^{\bullet}$** along with two monomer structures extracted from the $D_2/D_3$ and $D_0/D_1$ minimum energy crossing interactions of the complex, corresponding to frames 1 and 100 in Reference 14. Nonorthogonal state interaction of multistate density functional theory (MSDFT-NOSI) and time-dependent density functional theory (TDDFT) are used with PBE0 functional and cc-pVDZ basis set. Ground state ($D_0$) geometries are optimized using PBE0/cc-pVDZ. All energies are given relative to the doublet ground state of the respective molecules.

| Optimized Geometry | MSDFT-NOSI | | | | | TDDFT |
|---|---|---|---|---|---|---|
| | $D_0$ | $D_1$ | $D_2$ | $D_3$ | $Q_1$ | $D_1$ |
| MPTZ$^{+\bullet}$ | 0.0 | 1.60 | 3.32 | 4.55 | 4.12 | 1.62 |
| TEMPO | 0.0 | 2.69 | 6.21 | 7.55 | 7.49 | 2.75 |
| **A-R$^{\bullet}$** | 0.0 | 2.68 | 2.79 | 3.90 | 3.12 | 2.75 |
| **A-R$^{\bullet}$**($D_2/D_3$) | 0.27 | 3.03 | 3.23 | 4.51 | 3.58 | 3.09 |
| **A-R$^{\bullet}$**($D_0/D_1$) | 0.86 | 2.79 | 3.50 | 4.07 | 3.16 | 2.96 |

On the performance of MSDFT for treating excited states of free radical species, we find that the computed first and second excitation energies for TEMPO and MPTZ$^{+\bullet}$ are in good accord with those from LR-TDDFT calculations, corresponding to transitions, respectively, from the doubly occupied Kohn-Sham orbital to the singly occupied orbital ($H \rightarrow S$) and from $S$ to the lowest unoccupied orbital ($S \rightarrow L$). We note that in TDDFT calculations, there are one or two lower excitation energies than the state dominantly of $S \rightarrow L$ character, due to a lack of doubly excited configurations. Therefore, we only listed the value corresponding to the $H \rightarrow L$ transition from LR-TDDFT calculations. Experimentally, the observed excitation energies of TEMPO range from 2.61 eV (in hexane) to 2.92 eV (in water), in adequate agreement with MSDFT and TDDFT results.[53] The absorption of $MPTZ^{+\bullet}$, a very stable cation radical species,[54] was observed to have a peak maximum of 516 nm (2.40 eV),[51] clearly significantly higher than the values corresponding to $H \rightarrow S$ transition from MSDFT and TDDFT calculations. The origin of this difference is unclear, but interestingly, there is an absorption at 725 nm (1.71 eV) after the initial excitation and charge transfer in the $MPTZ^{+\bullet}\text{-}A^{-\bullet}R^{\bullet*}$ complex,[13] which is close to the computed $MPTZ^{+\bullet}$ excitation.

For the electron acceptor radical, **A-R**$^{\bullet}$, we determined the transition energies at three structures, the optimized ground-state geometry and the two monomer geometries in the MECI complexes. The first three doublet excited states ($D_1$, $D_2$ and $D_3$) have energies of 2.68, 2.79 and 3.90 eV from MSDFT (Table 2), corresponding to the single electron transition $H \rightarrow S$ to yield $D_1$ and a mixture between $S \rightarrow L$ and the multiconfigurational $H \rightarrow L$ transition for the latter two states. LR-TDDFT yields a reasonable value for the $H \rightarrow S$ transition as the second excited state which is listed in Table 2, but it failed to describe the multiconfigurational states with double excitation, placing the $S \rightarrow L$ transition as the first excited state at 1.93 eV. **A-R**$^{\bullet}$ absorbs at 524 nm (2.37 eV) and a component of 2.58 eV (480 nm) was assigned to this molecule in the donor-acceptor complex,[13] which are close to the first excitation energy of the free radical species from MSDFT calculations. The variations of the ground-state ($D_0$) and excited state with different geometries in the molecular complex appear to be reasonable in Table 2, but the excited states are mixed with multiconfigurational states involving $H \rightarrow L$ transition.

*5.2. Tripdoublet states of bimolecular exciplex.*

Song recently reported a study of the potential energy curves of the excited-state complex between MPTZ and the free radical acceptor **A**-**R**• using a spin-free formulation of XMS-CASPT2.[14] To illustrate the performance and capabilities of multistate density functional theory, we follow that work and compare the energies and qualitative trends between the two methods on structures along an interpolated path connecting $D_2/D_3$ and $D_0/D_1$ MECI points that were optimized using SA-CASSCF[3e,4o]/6-31G(d) coupled with a PCM model.[14]

We have not yet implemented a continuum solvation method in MSDFT calculations. Thus, we made a simple correction to our gas phase results by assuming that different neutral and ionic states have the same solvent effects, respectively. We used the IEFPCM model in Gaussian16 to determine the solvation energies for the ground state and a model ion-pair configuration, and the results are listed in Table 3. It turns out that the PCM solvation (free) energies are nearly unchanged for the "local" complex at different structures ($\Delta\Delta G_{sol} < 0.02\ eV$), but solvation has large effects on the charge-transfer states (Table 3). To model the ion-pair species, we protonated at the sulfur site of MPTZ to create an ion pair complex, which has a differential solvation free energy relative to the neutral complex ranging from -0.43 eV at the $D_2/D_3$ intersection (frame 1) to 1.03 eV at the $D_0/D_1$ crossing point (frame 100). Thus, solvation effects primarily affect the relative energies of the ionic states. We added the solvation differences (Table 3) to the MSDFT results (Table 4), and the shifted ionic states are included in Figure 6, shown in grey-colored dash curves.

**Table 3**. Estimated solvation energies (eV) for the the MPTZ- **A-R**$^\bullet$ complex in the ground state ($D_0$) and a sulfur-protonated ion-pair model. Polarizable continuum model mimicking methanol solvent was used along with PBE0/cc-pVDZ calculations.

| complex | $D_2/D_3$ (1) | 16 | 32 | 56 | 80 | $D_0/D_1$ (100) |
|---|---|---|---|---|---|---|
| neutral ($D_0$) | -0.54 | -0.53 | -0.54 | -0.54 | -0.55 | -0.55 |
| ion pair | -0.98 | -1.09 | -1.22 | -1.40 | -1.51 | -1.58 |
| $\Delta\Delta E_{sol}$ | -0.43 | -0.56 | -0.68 | -0.85 | -0.96 | -1.03 |

**Table 4**. Computed energies (eV) from MSDFT/PBE0/cc-pVDZ calculations. All energies are given relative to the ground-state energy at the $D_2/D_3$ crossing geometry (1) along an interpolated pathway at percent intervals 16, 32, 56, and 80, leading to the $D_0/D_1$ crossing geometry (100). Solvation effects on charge-transfer states are given, but they do not affect local states (neutral) appreciably (Table 3) and they are not listed.

| State | Transition | 1 | 16 | 32 | 56 | 80 | 100 |
|---|---|---|---|---|---|---|---|
| $D_0$ | **D-AR**$^\bullet$ | 0.00 | -0.23 | -0.31 | -0.21 | -0.01 | 0.46 |
| $D_1$ | **D$^{+\bullet}$-AR$^-$** | 2.70 | 2.32 | 1.98 | 1.76 | 1.60 | 1.78 |
| $D_2$ | **D$^{+\bullet}$-A$^-$$^\bullet$R$^{\bullet*}$** | 2.84 | 2.39 | 2.04 | 1.76 | 1.61 | 1.82 |
| $D_3$ | **D$^*$-AR**$^\bullet$ | 2.39 | 2.23 | 2.23 | 2.32 | 2.37 | 2.70 |
| $D_4$ | **D-AR$^{\bullet*}$** | 2.76 | 2.53 | 2.45 | 2.56 | 2.75 | 3.20 |
| $D_5$ | **D-AR$^{\bullet*}$** | 3.13 | 2.94 | 2.51 | 2.49 | 2.20 | 2.63 |
| $Q_1$ | **D$^{+•}$-A$^-$•R$^{•*}$** | 2.84 | 2.39 | 2.05 | 1.76 | 1.61 | 1.82 |
| $Q_2$ | **D$^*$-AR$^•$** | 2.39 | 2.26 | 2.25 | 2.32 | 2.37 | 2.70 |
| $D_1$(sol) | **D$^{+•}$-AR$^-$** | 2.26 | 1.76 | 1.29 | 0.91 | 0.64 | 0.75 |
| $D_2$(sol) | **D$^{+•}$-A$^-$•R$^{•*}$** | 2.41 | 1.84 | 1.36 | 0.91 | 0.65 | 0.79 |
| $Q_1$(sol) | **D$^{+•}$-A$^-$•R$^{•*}$** | 2.40 | 1.83 | 1.36 | 0.91 | 0.65 | 0.79 |

As in the work of Song,[14] all energies are given relative to the ground state at the $D_2/D_3$ crossing structure (Table 4). To make comparison convenient, we have kept the order of states in methanol solution to label states from the present MSDFT calculations, but we placed the curve-labels for excited states in the order of gas-phase results at the frame-1 structure in Figure 6. Without solvation effects, the first excited state ($D_3$) is the "tripdoublet" at 2.39 eV originating from the locally excited triplet state of MPTZ in the presence of the **A-R$^•$** radical doublet (the **D$^*$-AR$^•$** configurations in Figure 3). This state is essentially degenerate with the corresponding quartet state ($Q_2$). Figure S6 of Reference 14 showed that the energies for these states from XMS-CASPT2 calculations are about 0.1 eV higher, and the qualitative trend of the doublet-quartet degeneracy is reproduced in both methods.[14] These two states ($D_3$ and $Q_2$) continue to increase in energy as the structures convert from the $D_2/D_3$ crossing point to the final $D_0/D_1$ geometry, reaching a relative energy of 2.70 eV. Interestingly, there is a remarkable difference in dynamic correlation correction by more than 2 eV between the two MECI structures as seen in Song's work in comparison with SA-CASSCF energies (Figure S6 of Reference 14).

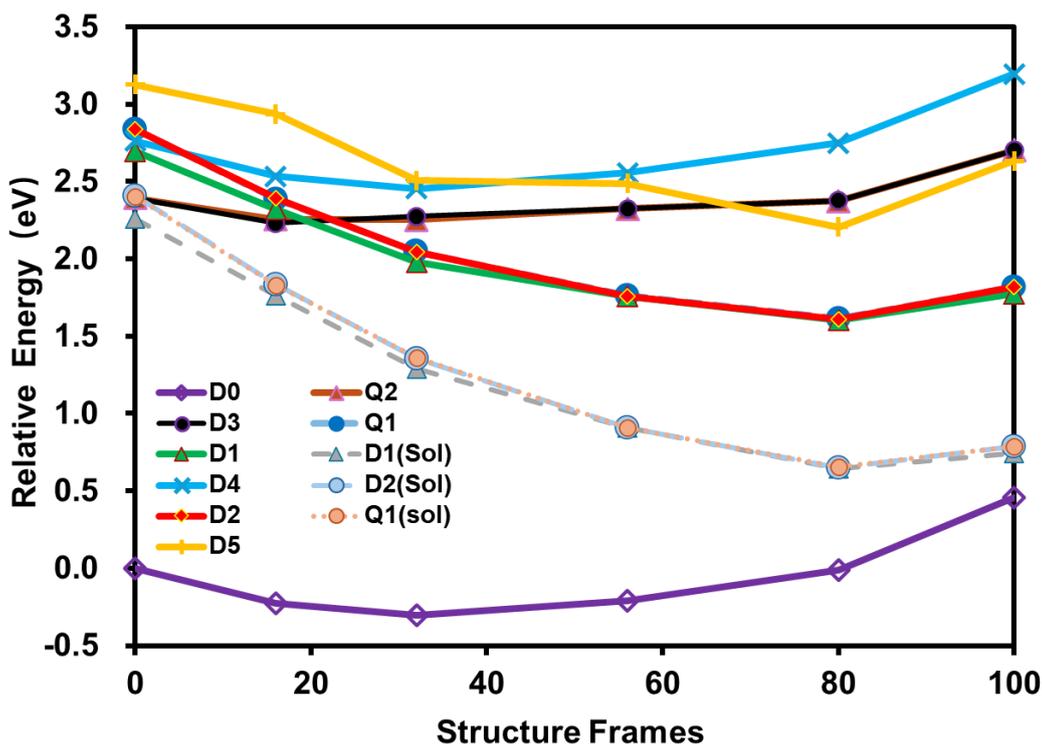

**Figure 6**. Ground and excited-state energies determined using nonorthogonal state interaction (NOSI) of MSDFT for the TMPZ(**D**)-**A**-**R**• radical complex along an interpolated pathway from the $D_2/D_3$ minimum energy crossing intersection (frame 1) to the $D_0/D_1$ crossing point (frame100). Solvation-corrected ion-pair states are displayed in grey-colored dash curves, which are lower in energy than valence excitations. States are labeled according to those in methanol solvent in the work of Song,[14] although the order in the figure legend is placed in the order for the complex at the frame 1 geometry in the gas phase. State $D_3$ (black) and $Q_2$ (maroon, behind) are local excitations of MPTZ, **D***-**AR**•, depicted in Figure 2. States $D_4$ and $D_5$ are transitions, respectively, from the highest doubly occupied (*H*) to the singly occupied orbital (*S*) and from *S* to the lowest unoccupied orbital (*L*) of the radical **A**-**R**•* in the presence of TMPZ ground state.

The MAS used in the present study also includes the $H \rightarrow S$ ($D_4$) and $S \rightarrow L$ ($D_5$) transitions of **A**-**R**•* local excitations (Figure 2). These states are generally higher in energy than other states and are not included in the work of Song, but $D_5$ emerges to be lower in energy than $D_3$ near the $D_0/D_1$ geometry from MSDFT calculations (Figure 6).

Turning to charge-transfer states, which was described to take place after the initial local excitation described in the original experimental work,[13] we found that solvation has differential and greater effects than on the covalent, locally excited states, ranging from -0.43 eV (-10.0 kcal/mol) at the $D_2/D_3$ geometry (frame 1) to -1.03 eV (-23.8 kcal/mol) at the other end of the path. Without solvation, MSDFT calculations place them above the MPTZ local excitations, but they quickly decrease to be lower in energy as the structure moves towards the $D_0/D_1$ crossing points. Coincidently, the solvated $D_2$ energy is degenerate (2.39 eV) with the $D_3$ states using MSDFT at the SA-CASSCF intersection geometry (frame 1), whereas there is a small gap between the two states from XMS-CASPT2 (Supplementary Figure S6 of Reference 14). Throughout the pathway, the energy curves are in close accord between MSDFT-NOSI and XMS-CASPT2 methods with a slight deviation of about 0.2 eV at the $D_0/D_1$ geometry, perhaps due to a very approximate correction to solvation effects in Figure 6.

A slight, but noticeable difference between XMS-CASPT2 and MSDFT results is that the former produces a nearly perfect energy degeneracy among three states, the direct charge transfer $D_1$ state (**D$^{+\bullet}$-AR$^-$** in Figure 3), the tripdoublet $D_2$ state and the quartet state ($Q_1$) of charge-transfer plus local excitation (**D$^{+\bullet}$-A$^{-\bullet}$R$^{\bullet *}$** in Figure 3).[14] In MSDFT-NOSI, the solvation-corrected (as well as that in the gas phase), the $D_1$ energy is somewhat lower than the tripdoublet $D_2$ state, with the largest difference at 0.15 eV near the $D_2/D_3$ geometry. MSDFT also revealed an essential energy degeneracy between the tripdoublet and its quartet counterpart as that from XMS-CASPT2 calculations.[14]

The observed magnetic field effects on product distribution have been attributed to different populations of the tripdoublet and quartet states, initially produced in the locally excited states ($D_3$ and $Q_2$), i.e., the **D**$^\bullet$-**AR**$^\bullet$ complex, and the charge-transfer tripdoublet and quartet states of the BRIP **D**$^{+\bullet}$-**A**$^{-\bullet}$**R**$^{\bullet*}$ ($D_2$ and $Q_1$).[13] It was suggested that **D**$^\bullet$-**AR**$^\bullet$ was first populated, which quickly relaxes via the $D_2/D_3$ conical intersection to the BRIP states, in addition to the direct charge transfer state **D**$^{+\bullet}$-**AR**$^-$ (Figure 3). The doublet charge-transfer states of BRIP can return to the ground state of the molecular complex. Local excited states lead to dissociation into neutral molecular products, whereas the charge-transfer states give rise to molecular ions, both of which depend on the applied external field and have been observed experimentally.[13] We have not modeled the interaction of different spin states with an applied magnetic field, which is not the main aim of this work. It would be interesting to carry out investigations into photodynamic processes in future studies.

## 5. Summary

Multistate Density Functional Theory (MSDFT) extends the traditional Hohenberg-Kohn Density Functional Theory (DFT) to encompass both ground and excited electronic states. By introducing a matrix density $\boldsymbol{D}(\boldsymbol{r})$ of rank $N$ as the fundamental variable, MSDFT establishes a one-to-one correspondence between $\boldsymbol{D}$ and the Hamiltonian matrix $\boldsymbol{\mathcal{H}}[\boldsymbol{D}]$. Minimizing the trace of the Hamiltonian matrix density functional, $E_{MS} = \min_{\boldsymbol{D}(\boldsymbol{r})} tr\{\boldsymbol{\mathcal{H}}[\boldsymbol{D}]\}$, directly yields the energies and densities of the $N$ lowest eigenstates. Notably, the matrix density $\boldsymbol{D}(\boldsymbol{r})$ can be exactly represented by no more than $N^2$ Slater determinants, preserving the computational efficiency inherent to Kohn-Sham DFT.

In this chapter, we present a nonorthogonal state interaction (NOSI) approach within the framework of multistate density functional theory to investigate systems characterized by triplet-doublet spin coupling interactions and open-shell photochemical reactions. Specifically, we examine the reaction between 10-methylphenothiazine and a dicarboximide electron acceptor linked to the stable free radical 2,2,6,6-tetramethylpiperidin-1-oxyl (TEMPO). This reaction produces a biradical radical-ion pair, with its yield influenced by the strength of an external magnetic field. Employing a minimal active space comprising ten Slater determinants, we achieve both qualitative trends and quantitative excited-state energies for local and charge-transfer doublet and quartet states. These results are in good accord with previous studies utilizing the spin-free formulation of extended multistate-complete-active-space second-order perturbation theory (XMS-CASPT2). Given the size of the exciplex in this study, MSDFT-NOSI demonstrates potential applicability to realistic molecular systems in light-emitting diodes and molecular spintronics.

**Acknowledgements**: This work has been partially supported by the National Natural Science Foundation of China (Grant nos. 12304285 and 22173107) and Shenzhen Bay Laboratory. Computational work carried out at Minnesota was supported by the National Institutes of Health (Grant GM046736).